# Big Data as a technology-to-think-with for Scientific Literacy

**Geovani Lopes Dias**
**Renato P. dos Santos**

**ABSTRACT**

This research aimed to identify indications of scientific literacy resulting from a didactic and investigative interaction with Google Trends Big Data software by first-year students from a high-school in Novo Hamburgo, Southern Brazil. Both teaching strategies and research interpretations lie on four theoretical backgrounds. Firstly, Bunge's epistemology, which provides a thorough characterization of Science that was central to our study. Secondly, the conceptual framework of scientific literacy of Fives et al. that makes our teaching focus precise and concise, as well as supports one of our methodological tool: the SLA (scientific literacy assessment). Thirdly, the "crowdledge" construct from dos Santos, which gives meaning to our study when as it makes the development of scientific literacy itself versatile for paying attention on sociotechnological and epistemological contemporary phenomena. Finally, the learning principles from Papert's Constructionism inspired our educational activities. Our educational actions consisted of students, divided into two classes, investigating phenomena chose by them. A triangulation process to integrate quantitative and qualitative methods on the assessments results was done. The experimental design consisted in post-tests only and the experimental variable was the way of access to the world. The experimental group interacted with the world using analyses of temporal and regional plots of interest of terms or topics searched on Google. The control class did 'placebo' interactions with the world through on-site observations of bryophytes, fungus or whatever in the schoolyard. As general results of our research, a constructionism environment based on Big Data analysis showed itself as a richer strategy to develop scientific literacy, compared to a free schoolyard exploration.

**Keywords:** Constructionism. Scientific literacy. Big data. High School. Crowdledge.

# Big Data como uma tecnologia-de-pensar-com para a literacia científica

**RESUMO**

Esta pesquisa teve como objetivo identificar indicações de alfabetização científica resultantes de uma interação didática e investigativa com o software de Big Data *Google Trends* por alunos do primeiro ano do ensino médio de Novo Hamburgo, no sul do Brasil. Tanto as estratégias de ensino como as interpretações de pesquisa se baseiam em quatro contextos teóricos. Em primeiro lugar, a epistemologia de Bunge, que fornece uma caracterização completa da Ciência que foi central para

**Geovani Lopes Dias** é Mestre em Ensino de Ciências. Atualmente, é professor de Física na rede pública estadual. Endereço para correspondência: R. Sarquiz Sarquiz, 49, apto. 101, Novo Hamburgo, RS.
E-mail: geovani.phy.dias@gmail.com.
**Renato P. dos Santos** é Doutor em Física. Atualmente, é Professor Adjunto do Programa de Pós-Graduação em Ensino de Ciências e Matemática - ULBRA. Universidade Luterana do Brasil/PPGECIM. Endereço para correspondência: ULBRA/PPGECIM, Av. Farroupilha, 8001, prédio 14, sala 338, 92450-900 Canoas, RS.
E-mail: renatopsantos@ulbra.edu.br.

o nosso estudo. Em segundo lugar, a estrutura conceitual de Alfabetização científica de Fives et al. que torna o nosso foco de ensino preciso e conciso, bem como apoia um dos nossos instrumento metodológico: o SLA (*Scientific Literacy Assessment*). Em terceiro lugar, o constructo 'crowdledge' de dos Santos, que dá sentido ao nosso estudo quando torna o desenvolvimento da alfabetização científica em si versátil para dar atenção aos fenômenos sociotecnológicos e epistemológicos contemporâneos. Finalmente, os princípios de aprendizagem do construtivismo de Papert inspiraram nossas atividades educacionais. Nossas ações educativas consistiram de alunos, divididos em duas turmas, investigando fenômenos escolhidos por eles. Foi feito um processo de triangulação para integrar métodos quantitativos e qualitativos nos resultados das avaliações. O delineamento experimental consistiu somente em pós-testes e a variável experimental foi a via de acesso ao mundo. O grupo experimental interagiu com o mundo usando análises de gráficos temporais e regionais de interesse de termos ou tópicos pesquisados no Google. A classe de controle fez interações 'placebo' com o mundo através de observações no local de briófitas, fungos ou qualquer outra coisa no pátio da escola. Como resultados gerais da nossa pesquisa, um ambiente construcionista baseado na análise Big Data mostrou-se como uma estratégia mais rica para desenvolver a alfabetização científica, em comparação com uma exploração livre do pátio de escola.

**Palavras-chave:** Construcionismo. Literacia científica. Big Data. Ensino médio. Crowdledge.

## INTRODUCTION

Our research builds its methodology and theoretical framework from the fact of the poor condition of literacy of Brazil, especially on scientific literacy, as primary reports of the area have pointed us. The data available show an awkward situation of education for a huge number of students (as well as the ordinary people). As an example, a high fraction (85%) of the Brazilian PISA population (15 years old students) is not able to interpret and use scientific concepts from different areas and apply them directly, to elaborate short statements using facts, or take decisions supported by scientific knowledge (OECD, 2013, p.113; BRASIL, 2013, p.50).

If this results was not bad enough, the skills to work in the technological arena that industry and market demands, including Industry 4.0, Internet of Things, that is to say, with the Big Data characteristics of the present world we live in, have not yet been acquired by enough people as to supply market demands, resulting in a gap between global market and network society (CASTELLS, 2005; EKBIA, et al., 2015; DOS SANTOS, 2014).

Given this context, how can teaching face this situation? Science teaching reviews point to two ways of developing Science literacy: an explicit and an implicit way, the difference lying in explicit (or not) heuristic principles, methodological or etymological, to guide students' thinking and doing. This (explicit) approach is recommended by reviews such as Lederman et al. (2013), Duschl & Grandy (2013), Tenreiro-Vieira & Vieira (2013), Fives et al. (2014), Marin & Halpern (2011), and Byrnes & Dunbar (2014).

The present work brings the results of a research with the goal of analyzing indications of scientific literacy arising from interaction with a teaching environment that relies on the technology of our network society: Big Data – or, as our construct already proposed, crowdledge: knowledge constructed-with-data-from-the-crowds (DOS SANTOS, 2015).

Our sample was composed of first-grade students from a small public high school in a city located in the southern region of Brazil. Our environment was inserted under the context of scientific literacy in Physics classes, whose teacher is one of the authors. The proposed activities were inspired by Papert's Constructionism (EDWARDS, 1998; PAPERT, 1980), as well as by the epistemological heuristic principles of Bunge (1989, 1998), and the scientific literacy framework of Fives et al. (2014).

## THEORETIC-METHODOLOGICAL FRAMEWORK

Here, a brief synopsis of the four epistemological pillars that support the methodological actions and cognitive judgments of this teaching research is presented.

### Bunge's epistemology of science

In his book *La ciencia: su método y su filosofía* (The Science: its method and philosophy, in a free translation), Mario Bunge paints the Science framework as a list of characteristics sitting on two epistemic columns: Science knowledge is necessarily rational, and its reality is objective.

The rationality of Science, argues Bunge, lies upon three bases. First, Science works with concepts (not images[1] or sensations, as e.g. art does), i.e., Science thinking begins and ends with ideas. Second, these ideas can be combined logically to structure other ideas (deductive inferences). Third, these ideas are not chaotically or temporally organised: there is systematicity in the way they are related, and the ordered sets are called theories.

The objectivity of Science, in turn, lies on the coherence between its objects, the concepts; this coherence being built by the verification whether those ideas adapt to the facts (e.g. measures). This conference in search of coherence is metaphorically described, in Bunge's work, as a controlled "trade" or "commerce."

Above these two aspects, Bunge builds the doing of Science, by some elements and heuristic rules that do not form a complete framework but provide a wide one. Given the limitations of space, the interested reader is referred to his book (BUNGE, 1998) or to our review (DIAS, 2016).

Meanwhile, how can Science be taught? Or how can one learn it? These questions interpret/translate the goal of this research. But still, there is the need for clarifying this concept: scientific literacy as a core framework for this learning.

---

[1] Images are used in Science only to build over rational underpinning, e.g., atomic models.

### Scientific literacy of Five et al.

Like many concepts in Science, especially in social Science, "scientific literacy" is under a thin and fragile consensus (LAUGKSCH, 2000; LEDERMAN et al., 2013). Among different visions, the solid literature research of Fives et al. (2014) was chosen. This choice was also grounded in our need of a research tool that, taking into account the Brazilian high-school characteristics, should not include:

- **Open/dissertative questions**, which would make it impracticable for classroom teachers, especially those at public high school, with a huge number of students and little time for them.

- **The requirement of specific scientific knowledge to answer it**, because if the class lacks it, the test will not identify literacy in general despite this specific knowledge.

- Scientific Literacy Assessment (SLA), the objective questionnaire built on that framework, carries a framework established in five[2] "components that together reflect [the authors'] perspective on the nature of scientific literacy" (FIVES et al. 2014). They are:

- **Role of Science**: is the understanding inherent to one's decision making, that concerns "(a) the kinds of questions that can be answered through Science, (b) the nature of scientific activities, and (c) generic scientific concepts present across field/discipline areas (e.g., variables, experiment, correlation, etc.)" (FIVES et al. 2014).

- **Scientific thinking and doing**: refers to skills (i.e., cognitive abilities) appropriately of Science, as to experiment, to observe with given criteria and to argue based on evidence.

- **Science, media, and society**: concerns to abilities as identify relations between Science and society, especially political and economic issues, as well as Science findings in media, in order to one acts as a critical citizen and consumer.

- **Mathematics in Science**: is the understanding of math needed to make functional those abilities inherent to the last component. For example, interpretation of charts and tables as well as geometric and statistical concepts in newspaper or advertising texts.

- **Science motivation and beliefs**: it is "three constructs [selected and] relevant to the successful engagement of scientific literacy": (a) subjective value of scientific tasks, (b) self-efficacy, in this kind of tasks (or how well one guess s/he do it), and (c) personal epistemology (or what one carries about the scientific knowing).

---

[2] Initially they listed six components, which were later collapsed into only five.

These components structure the framework that supports what in this research was called "scientific literacy." Certainly, it has limitations, as it do not include "abilities to interact with each other as they engage in scientific inquiry" (FIVES et al., 2014). However, as those authors point out, the framework was built for building a paper-and-pencil test; therefore, this goal has limited the coverage of the framework.

After some clarification of "Science" and "scientific literacy" in the context of this research, its other theoretical basis will be presented: Papert's constructionism, as well as how it was proposed to develop scientific literacy, namely by student interaction with Big Data.

### Papert's constructionism

Despite the sensorimotor and formal nature that human learning presents in Piaget's work (on which Papert has built his own),[3] Papert, based on his observation among various learning scenarios (scholar and ordinary ones), proposed the concrete thought as the central act by which someone learns something (PAPERT, 1993). If learning is progressively internalized actions, as Piaget's work proposed, to conceive the concrete thought as "principal" means that it "happens especially felicitously in a context where the learner is consciously engaged in constructing a public entity, whether it's a sand castle on the beach or a theory of the universe" (PAPERT; HAREL, 1991).

Some aspects of this assumption underpin the human learning itself, according to Papert's theory. They are: "felicitousness," "context," "learner," "consciously engagement," and "public entity."

As the center in the learning phenomena is the learner, this implies conceiving his/her thinking as, at least, **peculiar**, therefore, conceiving it **free** as well. However, the results of Papert's study suggested him that the learner thinking is **contextual** (i.e., depends on the whole condition available to thinking), because it is also **connective** (i.e., is empowered by what one knows), and **interactive** (i.e., is configured by operations – or "actions on objects" – yhat one has available to think). Because the learner thinking is peculiar, contextual, connective and interactive, he/she will learn better **freely** (i.e., having respected his/her own peculiarity), **concretely** (i.e., aiming and/or caring for exposition to others), **connectively** and **contextually** (i.e., having some availableness to enrich what is already known).

Connectivity and contextuality are well reached when the learner is in a **media** where he/she can find ways (preferably by him/herself) to act seeking his/her passion/s over which (at his/her own speed and on familiar issues) his/her thinking will progressively be more involved and/or effective in its objectiveness (to use an epistemological term). That media may be objects, software, researches, and so on. Nevertheless, given the complexity and singularity that one's learning demands, media will do it better the bigger the versatility it carries. So, learning environment must be structured in view of these elements.

---

[3] Due to our limited space here and the central role of Piaget's concrete thought in Papert's work, we do not discuss the sensorimotor and formal ones. The interested reader is referred to (PIAGET, 1993).

Indeed, the synopsis of Papert's Constructionism presented here can be more precisely described by seven implicit principles, extracted from Papert's work (PAPERT, 1980, 1993) and included in the master thesis (DIAS, 2016). They were named the **principle of human singularity**, the **principle of mathetic singularity**, the **psycho-thermodynamic principle**, the **principle of personal enculturation**, the **environmental-contextual principle**, the **principle of exteriorization**, and the **principle of bricolage.**[4] Given the limited space of this paper, however, they will not be discussed here. The interested reader is referred to the master thesis (DIAS, 2016).

Finally, this research, carrying the Papertian constructionism model of learning, asks for an environment that simulates scientific method, i.e., scientific thinking/doing. As pointed out (DOS SANTOS, 2014), and the results also confirms, a rich environment for the scientific method, especially for individuals in the society of our time, is the context of a Big Data analysis or a Big Data interaction. The one used is presented hereafter. First, due to its massive media bias, it is intended to clarify what Big Data is.

## BIG DATA: AN ICON OF THE TECHNOLOGY OF INTELLIGENCE OF THE NETWORK SOCIETY

Despite all definitions of Big Data related to its "bigness" (i.e., computer analysis involving massive volume, flow, variety, complexity – and so on – of its dataset, as many authors have already written, see e.g., HURWITZ et al., 2013, p.16), these authors understand its essential attribute is its analytic results as **weak emergence** (IBM, 2011; DOS SANTOS, 2014).

Two statements from Bedau may clarify this point: An "[…] emergent macro-phenomena somehow both depend on and are autonomous from micro- phenomena". Furthermore, "a macro-property of some system [...] is weakly emergent if and only if [this macro-property] is **generatively explainable** from all of the system's prior micro-facts but only in an **incompressible way**" (BEDAU, 2008, emphasis added).

An explanation is 'generative' if it explains, precisely and correctly, how macro-events happens in the timeline – like Newtonian Mechanics do to planetary movement. In turn, an explanation is called 'incompressible' when there are no shortcuts (in a valid, accurate and complete form) for generative explanations.

However, does Big Data make the human cognition dependent, in a network society context? The relevance of this issue lies, especially, on the enthusiasm identifiable on media about Big Data. After all, promising technologies are historical phenomena, as the press, the calculator, and the computer itself (POTS, 2014). Therefore, that "dependence relation" between individual and machine needs enlightenment.

---

[4] Papert and Sherry Turkle have adapted from anthropologist Claude Lévi-Strauss the use of this French word whose nearest (but inadequate) translation might be 'tinkering, – in the same way as the old-fashioned tinker would try one after another tool from his bag to fix the problem at hand, without being ever being upset in the slightest by its lack of the generality – the single, universally correct method that will work for all problems and for all people – so characteristic of our present school math and modern science (PAPERT, 1993, p.131, 143 – 144).

Pierre Lévy (1993) classifies the occidental society in at least three eras. The orality age, marked by allegories to keep oral society's knowledge alive, made myths a kind of antique dataset, ephemeral by individual memories. The writing age, characterized by copy and translation, "immortalized the voice" of the dead; somehow, people from this society tend to think categorically, while those from orality do it situationally. The press age gave birth to Science: press books had a speech structure different from mere old "monk writing," still marked by narratives.

At last, there is the society of computer (LÉVY, 1993) – or a network society, as described above (CASTELLS, 2005). Its novelty lies on simulation, which is an external and supplemental module to the imagination. What society one has (or will have) due to Big Data, given its characteristics? What learning may one get with it? This last issue is at the core of our research: what kind of Science learning a Big Data interaction (if possible) may provide? A proposal has made to answer it.

Dos Santos suggests a construct for Big Data (2015), **crowdledge**: emergent know**ledge** from **crowds**, i.e., knowledge provided by crowds' data and/or information spontaneously left (as digital footsteps and tweets). Hence, at the Big Data core is not only its bigness (JACOBS, 2009; EKBIA et al., 2015), as well as the emergent nature of its data.

As an example, some software may run algorithms as those used in Big Data. One of them, *Google Trends*,[5] do it by charts plotted based on Google historical search dataset: the users enter a search term, the software answer with the plotting. Many ideas, curiosities or concerns may occur to the curve plotting observer, even informally, especially when one enter more than a term and compare their curves among time.

Suppose one enters in Google Trends a term referring to a subject or topic that one likes (music style, sport, food, or whatever). Observing, on the charts, the behaviour of the total number of searches over a given time period, regionally or not, the user could perceive patterns and/or instigating behaviours that arouse the interest by its cause (on newspapers, journals, magazines, or even other Google Trends charts). This "movement" in search for causation, motivated by a personal love – made possible only by technology – is an example of how crowdledge can serve to a prototype experience of scientific research. In other words, by allowing individuality and interests to a crowdledge computational tool, people can immerse into a constructionist media of scientific method in Big Data software.

However, as this work carries a research inserted in a scholar context, a curriculum is required, what, however, instantly mischaracterize the research done as a genuine

---

[5] Accessible at: <http://www.google.com/trends>.

Papertian learning environment. In the methodology section, it is shown how this mischaracterization has been eased.

## GOOGLE TRENDS

The environment for literacy development was structured by using inquisitive interaction with Google Trends (GT) and is specified in the methodological section below. This section brings a presentation of GT.

The software of Google Inc. is mainly known by Google Flu Trends (GINSBERG, 2009), a particular extension of it. The software makes possible to users (even to those who haven't a Google account) accessing charts which are plotted based on Google Search dataset, as well as trends – up to a year in the future (GOOGLE, 2015). Another software, of the same type, is Google Correlate, which was not presented here; interested readers are referred to (DOS SANTOS, 2014; DIAS et al., 2015; MOHEBBI et al., 2011; BARAM-TSABARI, SEGEV, 2009).

GT works on an algorithm that correlates Google Search data to up to five terms inserted by the user to plot (up to) three-dimensional charts: the dimensions being time, relative and normalized interest score (RNIS), and colour (to distinguish the data terms). There are three kinds of charts, namely: time, regional and mean ones.

The time charts cross RNIS data of different terms. Here it is evident why RNIS are relative: they do not refer to the absolute volume of data. Rather, they are calculated relative to all searches put on Google (in each period, region and dataset – as news or shopping). In time-charts it is also clear why it is normalized: for it is calculated relative to the data on the chart. For example, the score "100" will ever be the highest one and other points is a value relative to it. Hence, RNIS' ordinate is not related to absolute topic's interest; instead, it refers to a relative quantity: both to all searches and to other items plotted. Figure 1 shows an example.

Figure 1 shows a comparison between three universities in the metropolitan region of Porto Alegre, in southern Brazil. The different colours represent each search terms; at horizontal axis is time (from 4/8/12 to 4/7/17); at the vertical axis is the interest score (from 0 to 100).

Regional charts, in turn, are two-dimensional chart (but it is possible to access a timeline option on it), namely, region (country, state, or city) clearly identified in maps, and region's RNIS, represented by colour intensity (as well as different colours to different topics). Figure 2 shows an example.

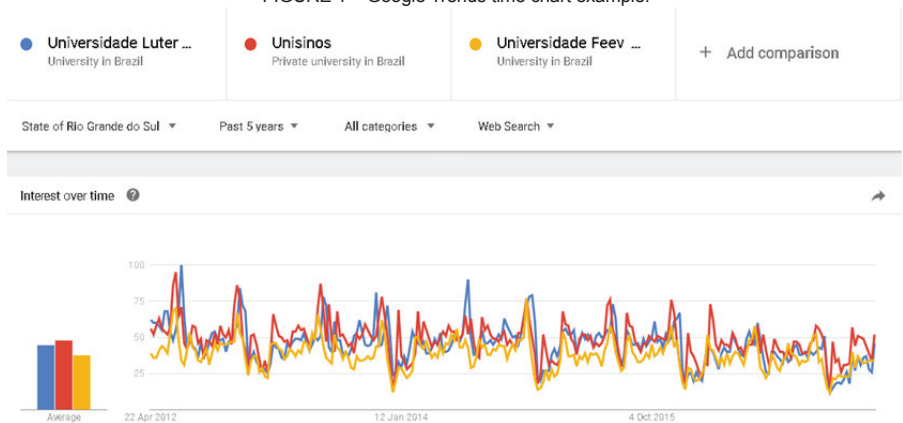

FIGURE 1 – Google Trends time chart example.

Source of data: Google Trends (http://www.google.com/trends).

On Figure 2, an example of interest by soccer in Brazil in the last year (from 22 Apr 2012 to 21 Apr 2017) is shown. On the left, Google Trends shows the map of the country (or the world, if the user chooses it) and, on the right, the scores by Brazilian state (sub-region, which implies countries, in the case of the world map).

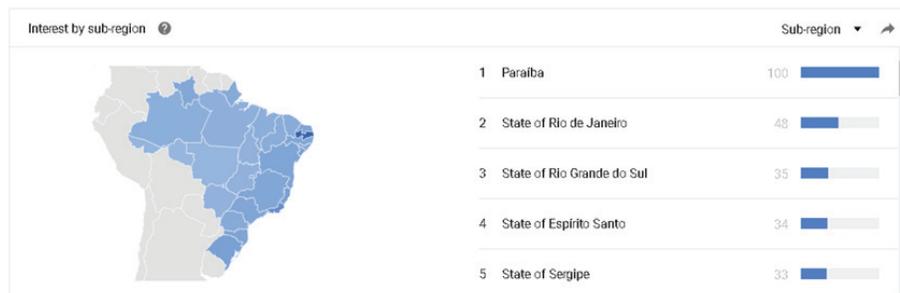

FIGURE 2 – Google Trends regional chart example.

Source of data: Google Trends (http://www.google.com/trends).

However, despite the considerations above, there are some inherent characteristics of those charts that may limit insights based on them. First, the time-dimension: it is not a "pure timeline," but also blurred by a hidden "technology accessibility" axis. As the access of particular population (e.g., see IBGE REF for Brazilian context) to Internet technologies grows (such as it happens with smartphones), more searches are made. That is to say, time-dimension has implicit to it an inseparable accessibility axis. Secondly, although the line-chart suggests otherwise, the points are *discrete* (at least by month – up to by minute, if there is enough dataflow). Thirdly, the RNIS-dimension (by its relative nature) carries a relative behaviour: i.e., a drop in the chart does not necessary means a decrease in the interest, but a drop in RNIS – moreover the absolute interest may have even grown, which is possible if the volume of all searches have increased more than those on the term.

The last considerations above are of great significance, especially to let students handle with GT and have more both precisely and coherent insights. Hereafter, it is described how GT was used as an environment intended to scientific literacy development.

## METHODOLOGY

Our research was a teaching experiment, whose object of analysis were two physics classes of a high school located at Novo Hamburgo, a metropolitan town in southern Brazil. The school, as well as the neighbourhood where it lies, is characterized by people in socioeconomic conditions diverse, but the majority lies on a low incoming range. It had, by the time of our, about 350 students. The two classes, in turn, differ mainly in its number of students and its shifts (morning and noon). Regarding other characteristics (as hereafter written), a pairing process was run. According to research ethics, terms of free consent were signed by the students' parents, besides an authorization from the school's principal.

The experiment was run coherently with the context of physics lessons classes. In order to amplify the understand of physics thinking and doing, as well as of Science generally, after a year of dialogical expositions about how to measure physical quantities as length, mass, force and temperature, i.e., teaching basic metrological knowledge and abilities as well as laws underpinning its measuring instruments (namely, single-pan balance, dynamometer and mercury-in-glass thermometer) – which composed the school's introductory physics framework – the classes had immersed in small and various investigations to develop basic concepts and abilities of Science research, as data, information, knowledge, correlation and construction of experimental tables and charts. Right after this immersive learning context, the experiences of our research have begun.

The experiment design chosen, underpinned on Campbell and Stanley revision, was the one **with experimental and control classes**[6] **and post-test only** (1966): with two sampling classes, one class was randomly selected as the control class; after the experiment run, the post-test was applied to both classes, as well as the students' reports were analysed, i.e., a mixed methods triangulation was made (KELLE; ERZBERGER, 2004).

According to Kelle and Erzberger (2004), the quantitative-qualitative scan in the empiric social research may mean a phenomenon being analysed by distinct methods or distinct aspects of a phenomenon (or distinct phenomena) to structure the clearest framework of reality. Therefore, triangulation is needed to amplify the scope, depth, and consistency of the research (FIELDING; FIELDING, 1986 quoted in FLICK, 1992).

---

[6] As students formed groups in activities, to avoid confusion, from here on, the experimental and control groups will be named CLASSES, and the students "teams", GROUPS.

As our research questions lie upon the characterization of Science learning with Big Data, the experimental variable was the "way (or lens) to read the world." The experimental class learned with Google Trends charts. The control class, in turn, learned with on-site observation at the schoolyard, the "lens" most commonly and broadly used in schools around the world (despite other available places).

## The control and experimental classes

As said, all the students in both classes had participated in our research, and both experimental class (EC) and control class (CC) had the same lessons[7] until the experimental and control classes have started. The experimental activity was designed to be very like the placebo activity, differing only by interaction with the world, as shown in Table 1: for EC via GT/GC and for CC by on-site observation.

TABLE 1 – Contrasting synopses of experimental and control classes.

| Activity | Experimental Class | | Control Class | |
|---|---|---|---|---|
| | *Activity's statements* | *Rules and guidance* | *Activity's statements* | *Rules and guidance* |
| 1 | Identification of an expression for a study on GT or GC. | The charts' behaviour can't be seasonal nor scholar | Identification of stuff on schoolyard of members' interest | To prefer details, the members never had seen. Take pictures of what was judged necessary. |
| 2 | Question development (and hypothesis, if chosen to do it) | None. | Question development (and hypothesis, if chosen to do it) | None. |
| 3 | Investigative actions planning (strategy) | Each student shall have activities to do | Investigative actions planning (strategy) | Each student shall have activities to do |
| 4 | Actions execution | Each student shall read at least one of the chosen texts | Actions execution | Each student shall read at least one of the chosen texts |
| 5 | Answer proposal to the questions developed | The charts observed on GT/GC shall be explained by the answer. The answer shall be set as definitive or not. | Answer proposal to the questions developed | The pictures taken shall be explained by the answer. The answer shall be set as definitive or not. |

---

[7] The physics content, i.e., its laws and concepts, were the context for these classes.

| Activity | Experimental Class | | Control Class | |
|---|---|---|---|---|
| | *Activity's statements* | *Rules and guidance* | *Activity's statements* | *Rules and guidance* |
| 6 | Report development and video recording | The charts, the questions, hypothesis, strategy, and results must appear on the report. The elucidation of interest on the chart behaviour, the clarifying of what was intent to explore and what was expected to find, as well as the indication of what and why each member does what he/she does. | Report development and video recording | The pictures, the questions, hypothesis, strategy, and results must appear on the report. The elucidation of interest on the stuff or details chosen, the clarifying of what was intent to explore and what was expected to find, as well as the indication of who do what and why each member does what he/she does. |

Source: Made by the authors.

In summary, the experimental class consisted of students inserting terms, from their personals interest, on GT or GC, until they see something that instigates their curiosity, like a behaviour of trends or a correlation chart (although none correlation was used). Over this, they were invited to follow the activities described in Table 1: write a/some research question/s, propose a strategy of investigation and execute it, suggest an answer grounded on data or information collected with their strategy and write a report to show it in a record. The themes of students' researches, as well as how they approached them, are presented in the Results section.

The control class, in turn, begins with students' free exploration at the schoolyard. The groups walked searching for something that, like EC, instigates their curiosity, which were mainly bryophytes (object chosen by students themselves) in all groups. Also like EC, the other activities listed in Table 1 were proposed.

## Limits of research validation

The external validation of our research is limited because one cannot assume that the classes are equal – in fact, no class at all can be –[8], despite the random chosen of the classes, as EC and CC, and the pairing process executed.

Furthermore, one should worry, above all, for scholar representativeness. Therefore, it was chosen to not take volunteers students on the counter-shift (Brazilian students have classes in only one shift) nor to divide one given class in control and experimental, since this would (in southern Brazilian high school context):

- **Internally invalidate** our research because students knowing they are executing different tasks would react differently, making their experience more artificial and biasing our teaching data.
- **Externally invalidate** it because untypical scholar experiences (either from class separation or extra class activities) would decrease the representativeness in the teaching field since they would be non-school ones.
- To make it **infeasible** because many students work or take courses on the counter-shift.

---

[8] Given the individual's cognitive singularity (a key-aspect of the Papertian constructionism) with which one thinks and acts in the world, we must admit that these singularities added to each other make all classes unique.

## Theoretical underpinning (and limits) of our learning environment

As well as validation limits, our work has theoretical limits that are mainly linked to our scholar context.

First of all, Bungean epistemology (1989, 1998) is used by essentially the following points of his heuristic scientific method, despite the many and various characteristics presented in his work:

a) *The need of scientific explorations for fact grounding.*

b) *The requirement for logical coherence between facts and ideas.*

c) *The facts transcendence in the Search by meaning and build of an artificial world (the scientific).*

d) *The need for verifiability, openness, clarity, and precision in inferences.*

e) *The both inquisitive and theoretical (to argue and to hypothesize) nature of Science.*

f) *The collective character of Science, due to partner cooperation as well as to other's knowledge in reports.*

These points are reached by the activities (which are also justified by these points) listed in Table 1 as well as their rules and guidance. So, the first activity (interaction with GT or schoolyard) is justified by points a, b and c; the second one, the inquisitive-thinking moment, when students were invited to build a questions that translate their curiosity, this activity is grounded by points b, d, and e; the third and fourth activities (strategy confection and execution) are underpinned in the point f; finally, the fifth one (student conclusion argumentation and reflection if they are, or not, definitive) is sustained in points b and d.

Secondly, Papert's Constructionism (1980, 1993) supports this environment by essentially five points: the care to individual singularity, the attention to the interactivity nature with which individual learning and thinking occur, the attention to the connectivity nature of those learning and thinking, the care to the contextual nature of those learning and thinking, and the need for concrete experience to the learning process.

Thirdly and finally, the learning environment also relied on Fives et al. (2014) Scientific Literacy framework to achieve theoretic-methodological coherence as well as on scientific teaching research. To do so, the following five points were especially a matter of concern: explicit and implicit classes articulation, the role of science achievement/development, science thinking and doing, mathematics knowledge, and motivations and beliefs of science.

Some of these Papert's Constructionism and Fives et al. (2014) Scientific Literacy framework aspects, however, were not fully contemplated by the GT environment. A full discussion of these issues can be found elsewhere (DIAS; DOS SANTOS, 2017).

## Quantitative data analysis

Our quantitative analysis consisted in running a t-test over the results of the Scientific Literacy Assessment, or SLA, from (FIVES et al., 2014). To do so, a pairing process was done, given that the number of students in each class was different as well as to build results external validation. The classes were paired via scholar means[9] and, when they were equal, via students' socioeconomic data (Table 2), which presented high correlation to scientific literacy in PISA (OECD, 2013).

TABLE 2 – Access to socioeconomic data.

| Information | Form of access |
| --- | --- |
| *Home computer | If available or not |
| *Internet connection | If available is or not |
| Suitable place to study | Whether it is satisfactorily silent according to the student, or not; if s/he has the necessary school supplies (i.e., notebook, writing supplies and calculator) or not. (The textbooks are provided by Brazilian Federal Government and, according to school's regulation, they must be at school); if s/he has her/his own courseware (as books, encyclopaedias, software, etc.) or not. |
| Familiar income | If it has its own vehicle (car and/or motorcycle) or not; if it has its own residence or not; how many smartechs (smartphones and/or tablets) there are at home; how many people are at home; the information marked with (*) above. |
| Parents schooling | The highest parent's schooling level:<br>- If 4th grade not accomplished: 0<br>- If 4th grade accomplished: 1<br>- If graduated from elementary school: 2<br>- If graduated from high school: 3<br>- If graduated from technologist course: 4<br>- If graduated from college: 5 |

Source: Made by the authors.

---

[9] As the high school scores were qualitative in this Brazilian state, we had to transform them to quantitative ones, which was possible due to those are only three grades. See how we made it in (AUTHOR1, 2016).

Some scores[10] were calculated individually to break the tie of scholar means on pairing process (not to superior inference making as a hypothesis).

After the pairing process and application of the SLA, to apply the t-test, the independence of observations and the normality and homogeneity of variance of samples was tested.

The independence of observations lies upon the uncommunicative culture of students, i.e., they did not meet each other (from different classes) with scholar purposes. Therefore, the literacy development of a class did not interfere with each other.

The normality was tested via Shapiro-Wilk test due to its power face to others tests and its reliability to small samples (THODE, 2002 quoted in GHASEMI; ZAHEDIASL, 2012). The skewness and kurtosis measurement were also calculated. All tests were done with the Charles Zaiontz's *Real Statistics Resource Pack* (Version 4.3)[11] add-in for Excel. The student scholar means by class as well as the results of normality test and measurement are presented in "Tabela 3" (Table 3) in our dissertation thesis. On Table 3 the normality results are shown.

TABLE 3 – Normality test and measurement results.

| Class | W Index | W Index Significance | Kurtosis | Kurtosis standard error | Skewness | Skewness standard error |
|---|---|---|---|---|---|---|
| Experimental | 0.9473 | 0.1840 | -1.01 | 0.87 | -0.31 | 0.45 |
| Control | 0.9210 | 0.1747 | 0.65 | 1.09 | -0.97 | 0.56 |

Source: Made by the authors.

Given the absolute value of the skewness and the absolute value of the kurtosis, as well as the given W Indexes, the classes' scores have a normal distribution; therefore, the t-test could be applied. As the normality tests, the t-test was done with the Real Statistics Resource Pack. We did so to the percentage of weighted arithmetical means and weighted harmonic means of SLA paired classes' scores.

## The Scientific Literacy Assessment

The SLA test is formed by two questionnaires which are underpinned by the previous Scientific Literacy framework: SLA-D (demonstrable) has two versions (SLA-

---

[10] To quantify these data, on every case but parents schooling and smartechs number, due to answers being binary (yes/no), they were quantified as 0 (zero) and 1 (one) for, respectively, no and yes. Then, some scores were calculated from unlike differences, namely: the sum of information "Silence", "School Supplies", "Own courseware," "Home Computer," and "Internet Connection," the sum of information "Home Computer," "Internet Connection," "Own Vehicle," "Own Residence," and "Smartechs per Capita"; the sum of all information.

[11] Copyright (2013 – 2016) Charles Zaiontz: <http://www.real-statistics.com>.

D1 and SLA-D2) with 19 multiple-choice questions each[12] and SLA-MB (motivations and beliefs), has 25 Likert questions.

SLA-D intends to access especially "demonstrable literacy," i.e., Applicable knowledge in the daily context. In turn, SLA-MB intends to access the student's view about Science knowledge and Science doing. Table 4 brings the categorized questions:

TABLE 4 – SLA question categorization.

| SLA | Literacy Elements | Questions | |
|---|---|---|---|
| | | D1 | D2 |
| D | Role of Science | 4,10,11,12,13,14,18 | 2,7,12,13,15,16,19 |
| | Scientific thinking and doing | 1,5,9,11,13,15,17,18 | 1,5,10,12,14,16,17,18 |
| | Science, society and media relations | 2,8,10,15,18 | 9,11,14,16,19 |
| | Math in Science | 3,6,7,9,12,16,19 | 3,4,6,8,9,10,13 |
| MB | Scientific knowledge worth | 20 to 25 | |
| | Science self-efficacy | 26 to 33 | |
| | Beliefs about Science | 34 to 44 | |

Source: Made by the authors based on (FIVES, 2014).

This test was translated into Portuguese by permission from its main author, Helenrose Fives (Email, 2015).

## Qualitative data analysis

Our qualitative vertex of triangulation was a content analysis of student reports; its goal was to identify evidences of scientific literacy in student's constructions, which leads to the *corpus*: their reports, posters, and records (only a few of the last ones were collected).

Specifically, the content analysis chosen was the propositional analysis of discourse (PAD). This technique, according to Bardin (2007), has inspiration in the structuralist hypothesis, which implies to (I) determine the themes (or core-references, CR) and to fragment the text in propositions, which surrounds the CR giving it the meanings. In other words, the CR are precisely the nucleus of the one's cognition about the message in his/her text (BARDIN, 2007).

To do so, two dimension types were established: the didactic dimension (DD) and the literacy dimensions (LD), all of them independent. The first type comes from activities proposed to students (Table 1). The second type comes from SLA elements testable by those reports. The DDs are shown in Table 5.

---

[12] A cut from 25 to 19 questions was made by Fives et al. themselves (2014) for a context like ours.

TABLE 5 – Didactic dimensions of the qualitative research branch.

| Dimension number | Dimension name | Description |
|---|---|---|
| 1 | Research questions (Q) | List of students' investigative questions/curiosities. |
| 2 | Hypothesis (H) | List of students' hypotheses that they, perhaps (it was optional), have proposed to themselves. |
| 3 | Coherence Q-H | The logical coherence between the research questions and the hypothesis, i.e., does the hypothesis really answer the questions asked? |
| 4 | Strategy adopted (S) | The students' methodological steps proposed by and to themselves. These were the ones they judge needed and feasible (given the time they had). |
| 5 | Coherence Q-S | Same as DD3, but between students' investigative questions and strategy. |
| 6 | Causal explanation (K) | The answer/s elaborated post-strategy execution for the posed questions. |
| 7 | Coherence Q-K | Idem DD3, but between students' investigative questions and causal explanation. |
| 8 | Chart analysis | The students' interpretations of charts (like GT ones) and/or tables. EC only exhibited it.[13] |

Source: Made by the author.

These dimensions were chosen according to the *corpus*. Given the interdependence of the reports, posters, and records, they were treated as a triune entity.

The qualitative dimensions DD and LD may be considered to intersect each other as follows. According to Fives et al. (2014), LD1 is structured by "Identify questions that can be answered through scientific investigation; understand the nature of scientific endeavours; understand generic Science terms/concepts." In this way, LD1 is assessed by DD's that assess it (1, 2, 3, 6): e.g., to elaborate questions and (possible) hypothesis and conclusions (questions' answers), coherently or not, students' groups show to understand/use right or wrong, generic scientific concepts, like "hypothesis" and "investigation", showing there their literacy (LD) by those activities (DD).

LD2, in turn, enclose the mastery of the following skills: "describe natural phenomena, recognize patterns; identify study variables; ask critical questions about study design; reach/evaluate conclusions based on evidence" (FIVES, 2014). These criteria are assessed by DD5 to DD8, when, e.g., students provide strategies or answers to their questions: by doing so, they articulate those skills, as "identify study variables" and "reach conclusions based on evidence."

---
[13] The CC, due to their chosen themes and investigative questions, did not work with charts or tables, as much their teacher had encouraged them.

Finally, LD3, which carries as skills "mathematics in Science [and] understand [its] application" (FIVES, 2014), as tables, charts, equations as well as math propositions, is assessed by DD8, where students demonstrate (or not) to interpret well GT charts according to the nature of the data (RNIS).

The intersections of the dimensions DD and LD are shown in Table 6.

TABLE 6 – Intersection of DD and LD.

| Dimension | Role of Science | Science thinking and doing | Math in Science |
|---|---|---|---|
| Research questions (Q) | X | | |
| Hypothesis (H) | X | | |
| Coherence Q-H | X | | |
| Strategy adopted (S) | X | X | |
| Coherence Q-S | | X | |
| Causal explanation (K) | X | X | |
| Coherence Q-K | | X | |
| Chart analysis | | X | X |

Source: Made by the author.

## RESULTS AND ANALYSIS

Here, results and analysis are presented, starting with the quantitative results, i.e., the t-test over SLA answers, followed by the PAD over the students' material.

### Mean scores tables

In Tables 7 and 8, we bring the averages of the SLA results from each of the classes, experimental and control. The t-test was done after the pairing process, which was done as described in the Methodology section (as the pairing and mean tables are too large, they were not included here; interested readers are reported to our master thesis (DIAS, 2016), "Tabelas" 5 to 9 and 12, as well as "Quadro" 16). The one-tail t-test was done with the Real Statistics[14] Excel add-in. In Table 7, we show the mean, deviation, as well as other values related to t-test. These Table discriminate results first for weighted mean, harmonic mean and secondly for the specific scores; the alpha value as well as its critical t-value and degrees of freedom (dg) are indicated.

---

[14] Real Statistics Resource Pack do programa MS Excel elaborada por Charles Zaiontz, (Versão 4.3). Copyright (2013 – 2015) Charles Zaiontz: <http://www.real-statistics.com>.

TABLE 7 – T test results for SLA-D (alpha = 5%; dg = 14; t-critical = 1.76131).

| Score | Weigh. mean[15] | | Harm. mean[16] | | Role | | TD | | SM | | Math | |
|---|---|---|---|---|---|---|---|---|---|---|---|---|
| Class | E | C | E | C | E | C | E | C | E | C | E | C |
| Mean | 0,40 | 0,40 | 0,33 | 0,34 | 0,40 | 0,40 | 0,35 | 0,35 | 0,51 | 0,47 | 0,36 | 0,36 |
| SD | 0,13 | 0,16 | 0,13 | 0,19 | 0,28 | 0,20 | 0,17 | 0,20 | 0,17 | 0,22 | 0,22 | 0,20 |
| t | 0,17 | | -0,19 | | 0 | | 0 | | 0,51 | | 0 | |
| p value | 0.43 | | 0.42 | | 0.50 | | 0.50 | | 0.31 | | 0.50 | |

Source: The authors, via Zaiontz's *Real Statistics Resource Pack* MS Excel add-in.

TABLE 8 – T test results for SLA-MB (alpha = 5%; dg = 14; t-critical = 1.76131).

| Score | Values | | AE | | Belief[17] | |
|---|---|---|---|---|---|---|
| Class | E | C | E | C | E | C |
| Mean | 0,72 | 0,74 | 0,57 | 0,57 | 0,46 | 0,59 |
| SD | 0,13 | 0,16 | 0,13 | 0,13 | 0,14 | 0,17 |
| t | -0,38 | | -0,08 | | -2,07 | |
| p value | 0.35 | | 0.47 | | 0.03 | |

Source: The authors, via Zaiontz's *Real Statistics Resource Pack* MS Excel add-in.

Tables 7 and 8 reveal that there was no difference between the classes, except on the element Beliefs, given the one-tail test p-values. These results gave us two insights. Firstly, our learning environment could help both classes reach the paired scores. Given the fact that EC hadn't some math classes, especially those to develop skills related to think-with charts and sheets, we can conclude that our environment really was helpful to them. Secondly, our activities could provide an environment which has developed a sceptical or even "less naive" sense about Science. Given the statistic difference between the classes and the EC lower score, which means SLA-MB answers more carefully about some extravagant proposition about Science, as "Whatever the teacher says in Science class is true" or "Once scientists have a result from an experiment that is the only answer", we can infer the less naive view about Science the environment could provide.

---

[15] Weighted arithmetical mean $\%\mu_a = \frac{1}{4}\sum_{i=1}^{4}\frac{p_{a,i}}{q_i}$, where $p_{a,i}$ is the student's score on the $i$-th literacy element, assessed by $q_i$ SLA-D questions.

[16] Weighted harmonic mean $\%H_a = \frac{1}{6{,}75}\frac{\sum_{i=1}^{4}q_i}{\sum_{i=1}^{4}\frac{q_i}{p_{a,i}}}$, where $p_{a,i}$ and $q_i$ are defined as above.

[17] Likert weighted arithmetical mean $\mu_{a,i}L = \frac{1}{q_i}\sum K_{a,i}$ where $K_{a,i}$ is the student's opinion (in Likert scale from 1 to 5) on the $i$-th" literacy element, assessed by $q_i$ SLA-MB question.

## Report reading inferences

Due to huge limitations on translating text data, as some meaning is lost in the use of the language, we refer the reader to our master thesis (DIAS, 2016) (in Portuguese) and present here just our inferences upon them as well as some examples to underpin them. The translating of proposition braking is accessible in "*Quadros*" (Charts) 17 to 28 in that dissertation.

A first inference is the hallmark feature of the experimental class was the variety of themes evoked to students' investigations. While the control class, restricted to the schoolyard, opted by bryophytes (moss, slimes, and trees), the experimental class chose a different theme for each group. This disparity certainty lies in the imagination borders offered to experimental class with the GT (i.e., any subject was allowed); after all, besides the bryophytes and architectural structures, what students are more expected to find in the desert (of people) schoolyard? At least, it seems that it was not enough to trigger their imagination in a plurality of themes.

A second inference was the "hypothesis-perplexity" exhibited by experimental class: they made no hypothesis throughout their researches. We understand that control class could propose a hypothesis, because it is simpler, due to they are common sense laden, to hypothesize over everyday objects or environments. What about GT charts? It seems this level of Big Data analysis was more complex and caused this "hypothesis-perplexity" given the data nature, unimaginable without the software. If a rupture with common sense is desired to promote Science thinking (as a cognitive instability for learning), a more perplex environment is desired.

A third inference was the lack of "conceptual debugging" resulting in the similarity between the "longed" scientific thinking and the (common sense) one presented by the apprentices after the interactions with the environment. No more elaborate investigative strategy was observed on any of the groups. It stems from the aforementioned fact that the learning environment (GT) lacks a feature that allowed for the apprentice to identify his conceptual mistakes by himself and learn from them.

A fourth inference was a kind of a "semantic boldness" presented by class E, with a greater proportion of groups exhibiting inferences than the class C, even if such inferences were extrapolations, without presenting additional sources that allowed them, or misinterpretations, that did not coincide with the nature of the data. They might stem from frustrating students' expectations about the limited "answers" provided by GT about the phenomena the students were interested in.

A fifth inference concerned objectivity. When analysing the Chart analysis DD, it was found that Class E was characterised by a greater number of pairs of coherence between the investigative inquisition and their cognitive judgments. It can be understood that a difference between the experimental and placebo environments is at the heart of this discrepancy in said number of pairs, a difference that refers to the more objective character of the GT crowdledge environment. In it, the world is accessible through graphics, so that this access depends, to some extent, on trivial graph reading ability. Therefore, the crowdledge environment offers more easily interpretable phenomena than those accessible by in situ observations (placebo environment), which depend to some extent on training and knowledge (Theory) on what is precisely intended to observe.

## CONCLUSIONS

As general results of our research, a constructionist environment based on Big Data analysis to develop scientific literacy showed a richer strategy for that development, compared to a free schoolyard exploration, and was statistically as effective as the control environment under the SLA framework.

Our judgments would only point to a possible trend in the human spirit along Big Data analysis. Although our conclusions may extend over our sample, our Big Data society, an Internet of Things society urges some questions, which our results may concern with some issues. For instance, what mental models' transformation would be expected in a Big Data society? What kind of epistemic and methodological Science knowledge would emerge in individuals who pass by Big Data interaction? How to catch this emergency?

It seems that our society is passing through changes we all might not be able to see them all. Perhaps, to use the context of our time, as our work did, to promote independent cognition to individuals, could be a way to a more equality society, where changes, in fact, are a natural law.

## ACKNOWLEDGEMENTS

We thank CAPES and FAPERGS for funding this research.